\documentclass[%
 aip,
 amsmath,amssymb,
 preprint,%
]{revtex4-1}

\usepackage{graphicx}
\usepackage{dcolumn}
\usepackage{bm}

\usepackage[utf8]{inputenc}
\usepackage[T1]{fontenc}
\usepackage{mathptmx}
\usepackage{etoolbox}

\usepackage{chemformula} 
\usepackage{hyperref}

\usepackage{comment}

\usepackage{xcolor}

\newcommand{\mbf}[1]{$\mathbf{#1}$}
\makeatletter
\def\@email#1#2{%
 \endgroup
 \patchcmd{\titleblock@produce}
  {\frontmatter@RRAPformat}
  {\frontmatter@RRAPformat{\produce@RRAP{*#1\href{mailto:#2}{#2}}}\frontmatter@RRAPformat}
  {}{}
}%
\makeatother
\begin{document}

\preprint{AIP/123-QED}

\title[]{Orientation Adaptive Minimal Learning Machine: Application to Thiolate-Protected Gold Nanoclusters and Gold-Thiolate Rings}
\author{Antti Pihlajam\"{a}ki}
\affiliation{Department of Physics, Nanoscience Center, University of Jyv\"{a}skyl\"{a}, FI-40014 Jyv\"{a}skyl\"{a}, Finland}

\author{Sami Malola}
\affiliation{Department of Physics, Nanoscience Center, University of Jyv\"{a}skyl\"{a}, FI-40014 Jyv\"{a}skyl\"{a}, Finland}

\author{Tommi K\"{a}rkk\"{a}inen}
\affiliation{Faculty of Information Technology, University of Jyv\"{a}skyl\"{a}, FI-40014 Jyv\"{a}skyl\"{a}, Finland}

\author{Hannu H\"{a}kkinen}
\email{hannu.j.hakkinen@jyu.fi}
\affiliation{Department of Physics, Nanoscience Center, University of Jyv\"{a}skyl\"{a}, FI-40014 Jyv\"{a}skyl\"{a}, Finland}
\affiliation{Department of Chemistry, Nanoscience Center, University of Jyv\"{a}skyl\"{a}, FI-40014 Jyv\"{a}skyl\"{a}, Finland}

\keywords{Machine learning, distance-based, force vectors, clusters, gold}

\date{\today}

\begin{abstract}
    Machine learning (ML) force fields are one of the most common applications of ML methods in the field of physical and chemical science. In the optimal case, they are able to reach accuracy close to the first principles methods with significantly lowered computational cost. However, often the training of the ML methods rely on full atomic structures alongside their potential energies, and applying the force information needs special modifications to standard algorithms. Here we apply distance-based ML methods to predict force norms and estimate the directions of the force vectors of the thiolate-protected gold nanoclusters. The method relies only on local structural information without energy evaluations. We apply the atomic ML forces on the structure optimization of the gold-thiolate rings, \ch{Au25(SCH3)18} nanocluster and two known structural isomers of the \ch{Au38(SCH3)24} nanocluster. The results demonstrate that the method is well-suited for the structural optimizations of the gold-thiolate systems, where the atomic bonding has a covalent nature in the ligand shell and at the metal-ligand interface. The methodology could be seen as an early attempt to introduce equivariant learning to distance-based ML methods.
\end{abstract}

\maketitle

\section{Introduction}

Monolayer-protected clusters (MPCs) are chemically diverse nanostructures consisting of metallic core, protecting organic ligand layer and an interface structure between \cite{tsukuda15}. The ligand layer stabilizes the metal particles, which would otherwise agglomerate or react with outside environment. Stabilization enables MPCs to have atomically precise structures. This chemically complex yet atomically well-defined nature of MPCs makes them an interesting research subject, where possible applications vary from catalysis and biological imaging to nanomedicine \cite{tsukuda15,malola21}. Understanding the operational mechanisms of the MPCs in these applications requires development of efficient and reliable novel computational strategies.

Density functional theory (DFT) was introduced over half a century ago by Hohenberg and Kohn \cite{hohenberg-kohn-64} and it has developed into the main tool in the field of computational nanoscience. However, DFT often requires lots of computational resources to be run in a reasonable amount of time. This has lead into development of various force fields, which accelerate the computations. For MPCs there have been developed, for example, ReaxFF \cite{bae13} and AMBER-GROMACS \cite{pohjolainen16} force fields. The drawback of these methods is that one has to compromise accuracy and often one still needs to do extensive parameter optimization. The introduction of machine learning (ML) methods to physical and chemical sciences have offered alternative approaches to atomic simulations. ML methods are not strictly bound by predefined mathematical functions imitating physical and chemical behavior but they are used to find underlying trends on given data. This has lead into numerous ML force fields, which are able to produce similar behavior of atoms as DFT in well-defined cases with fewer computational resources \cite{noe20,unke21,friederich21}. However, even if ML force fields are one of the most common applications of ML methods in the research field, underlying algorithms are general and their applications are not restricted on force fields. They also have many application in material informatics \cite{schmidt19,schleder19}, catalysis research \cite{toyao20} and they can even be trained to build materials \cite{ASLA19,ASLA20-1,ASLA20-2}.

\begin{figure}
    \centering
    \includegraphics[width=0.5\columnwidth]{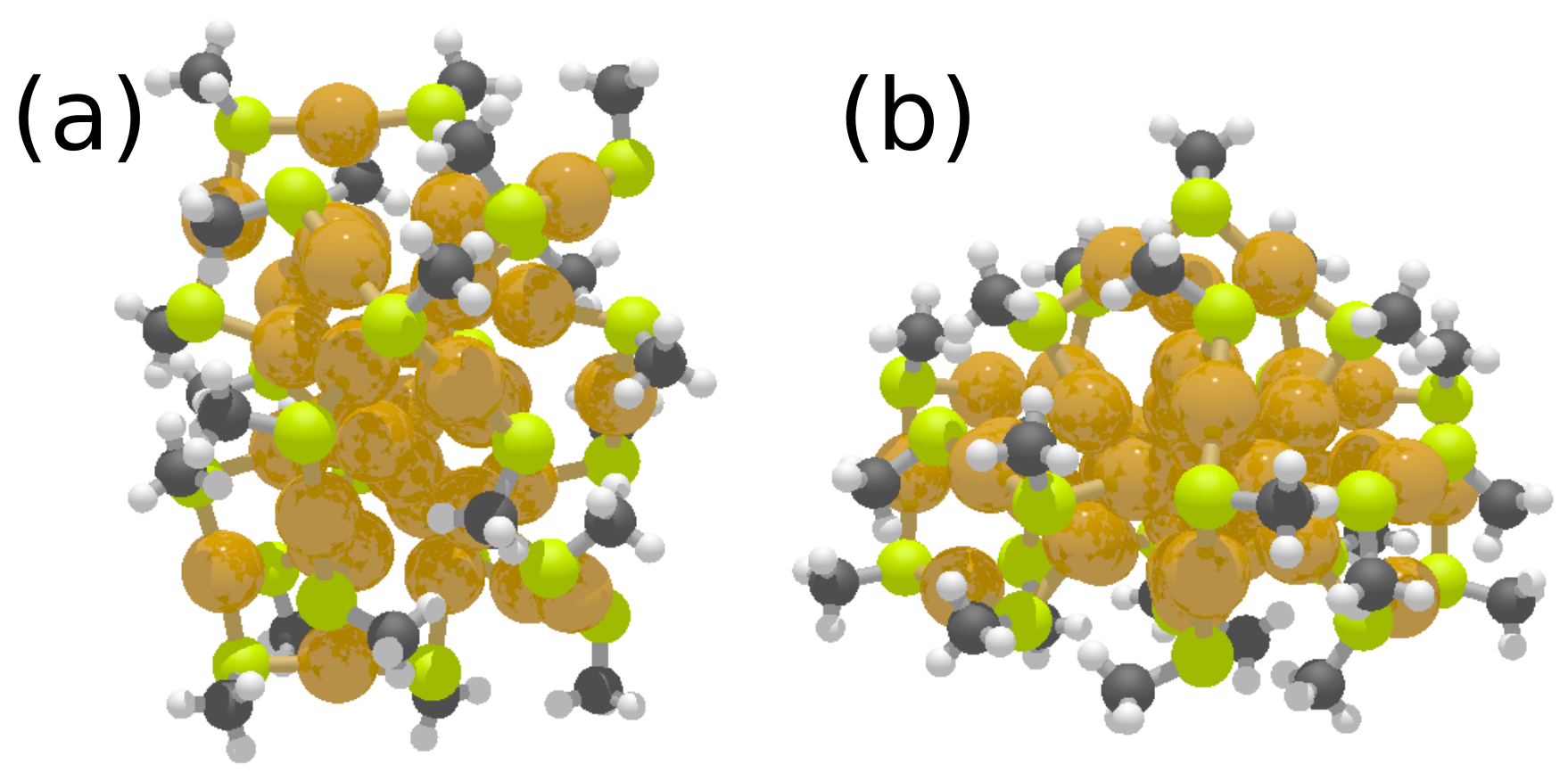}
    \caption{Two structural isomers of the \ch{Au38(SCH3)24} nanocluster: (a) the Q isomer \cite{qian10} and (b) the T isomer \cite{tian15}. The structures consist of metallic 23 gold atom core and protecting ligand layer. Metal-ligand interface is constructed from $[$Au-\ch{SCH3}$]_x$ oligomers or units of various lengths. Colors: orange, gold; yellow, sulfur; gray, carbon; white, hydrogen.}
    \label{fig:orig_struc}
\end{figure}

MPCs form a challenging nanomaterial class for ML methods, because of their chemical complexity and general low-symmetry molecular structure. However, there have been some successful studies on the subject. For example, artificial neural networks and support vector machine have been used to study synthesis and properties of MPCs \cite{wang19,copp20}, a rule-based method has been utilized to compare local atomic environments and to construct metal-ligand interfaces \cite{malola19}, and distance-based ML methods have been used to predict potential energies of \ch{Au38(SCH3)24} nanocluster for finite temperature Monte Carlo simulations of their dynamical properties\cite{pihlajamaki20}. \ch{Au38(SCH3)24} is also the focus of this study. This MPC has two known isomers: a cylindrical Q isomer \cite{qian10} and an oblate-like T \cite{tian15}. The structures are visualized in FIG. \ref{fig:orig_struc}. 

The structural difference of these two isomers can be highlighted by writing their chemical formula using the "divide and protect" idea \cite{hakkinen06}.  This means that the metallic core and protecting layer can be thought as separate entities and naturally notation should emphasize it. 
This way Q isomer could be written as 
Au$_{23}$@$[$SR-Au-SR-Au-SR$]_6[$SR-Au-SR$]_3$ and T isomer 
Au$_{23}$@$[$SR-Au-SR-Au-SR-Au-SR$]_2[$SR-Au-SR-Au-SR$]_3[$SR-Au-SR$]_3[$SR$]^{b}_1$, 
where the superscript $b$ refers to a bridge site and R denotes the organic part of the thiolate. In this notation it is clear that both isomers have 23 gold atom core and protecting layers consisting of gold-thiolate oligomers or units of varying lengths. Both isomers have been found experimentally and Q isomer is thermodynamically more stable than T isomer as shown both by experiments and DFT calculations \cite{tian15,rosalba19,taylor17}. Having two distinct structural isomers makes this MPC a very appealing testing ground for ML methods, because one can use the data from both isomers to test the generalizability of the method.

In this study we present a local force-based ML approach to simulate atomic systems. There are already some ML methods that can be trained with forces only, such as GAP \cite{GAP,GAP-rev}, sGDML \cite{sGDML}, NequIP \cite{nequip} and PaiNN \cite{painn}. However, methods like GAP and sGDML rely on explicit derivatives applied to the kernel matrix. Derivatives can be costly to calculate, if analytic forms are not available. Sometimes calculation of the analytic derivatives is not feasible for variety of reasons. This might be because of the complexity of the mathematical expression but more often it is simply due to the practical reasons. One might be using for some part of the method a separate pre-made package, which does not have derivatives. In order to generate correct derivatives one would need to fully understand how the base method is implemented. In this case user might lack the resources to further explore technical details or the access to the source code might be limited by a commercial license.

Equivariant learning methods, such as NequIP \cite{nequip} and PaiNN \cite{painn}, address this in a different manner. They learn representations directly from the atomic numbers and coordinates in the spirit of SchNet framework \cite{schnet1,schnet2,schnetpack}.  Instead of just forming invariant representations, which are adequate to predict scalar values such as potential energy, they also preserve orientation information. This enables the prediction of directional output, such as force vectors and stress tensors. However, this kind of equivariant learning methods are currently relying on complex deep neural networks (NNs). Conventional kernel methods are still mostly utilizing derivatives.

Instead of training a ML method to predict potential energies for given configurations and then taking a gradient to obtain forces, we train our method to predict directly force vectors subjecting to individual atoms. According to the Hellman-Feynman theorem, if the  Born-Oppenheimer approximation is valid, the forces are true quantum mechanical observables \cite{hellman37,feynman39}. Hence, they can be solved analytically separately from the energy calculation, which justifies the approach to use ML to predict forces directly. The goal is to create a model that handles atoms locally, which gives it a great potential to be generalized over different systems with similar local features. This kind of an generalizability has shown to be achievable at least for methods predicting electron density \cite{fabrizio19, grisafi19}. 

Estimating force vector directly independent from the energy has some special advantages. Forces are local properties of an atomic system, therefore the developed ML method would naturally be local and have promising generalization possibilities. Many methods are designed to predict how much every atom contributes to the potential energy \cite{GAP,GAP-rev,schutt17,schnet1,schnet2,schnetpack,chen18,nequip,painn,spookynet} and it has become standard approach in many ML applications. However, training a model with energies is still dependent on full structures of the atomic systems, because potential energy cannot be unambiguously separated from the full structure. Hence, the ML method has to learn on its own to divide the energy into local contributions, which is not a trivial task. On contrary, the model focusing solely on forces is not constrained by the use of full atomic structures. One could freely collect data sets from various systems with similar chemical composition, which eases the data generation process. The disadvantage of this approach is that forces are not strictly energy conserving anymore, therefore one has to consider what applications are suitable for this kind of methods.

There have been attempts to predict directly force vectors from atomic data of metal nanoparticles and surfaces \cite{botu15,botu17,pattnaik20}. However, these attempts use rotation variant representations of atomic environments instead of conventional rotation, translation and permutation invariant descriptors. This enables one to use conventional machine learning tools but introduces a new drawback: one has to somehow cover the orientation space. This is still a viable for lattice based systems with high symmetry. For low symmetry systems, this kind of an approach requires alignment of atomic environments and/or large amounts of rotated data. In order address this issue, the developed ML method would have to somehow utilize the idea of equivariance as mentioned above.

Our method uses conventional invariant descriptors and the ML method itself is made orientation adaptive. The approach enables fair comparison of chemical environments as the commonly used descriptors, such as Smooth Overlap of Atomic Positions (SOAP) \cite{bartok13}, Atom-Centered Symmetry Functions (ACSF) \cite{behler11}, Many-Body Tensor Representation (MBTR) \cite{huo17} and numerous other descriptors \cite{musil21}, are already well-known and tested. Our method breaks the force prediction task into two parts: (i) prediction of the norm of the force and (ii) estimation of the direction. Both parts utilize the so-called distance-based ML, which also enables elegant prediction of different attributes from the same similarity matrix. The similarity measure, as the name suggests, is the Euclidean distance. Our framework could be considered as a proto-equivariant method, because it uses invariant representations to predict properties and by orientation adaptivity it addresses the possible rotations of the system. It is an early attempt to implement equivariance into distance-based ML methods.

We trained and tested the method by using the data previously generated from DFT-level molecular dynamics (MD) simulations of the two structural isomers of \ch{Au38(SCH3)24} nanocluster \cite{rosalba19}. This data has already been used to predict potential energies using distance-based ML \cite{pihlajamaki20}, therefore this study provides a logical continuation to the previous research. We have tested extensively different parameters related to the method and applied it to the structure optimization of four different systems: gold-thiolate rings, \ch{Au38(SCH3)24} with outstretched protecting units in its ligand shell and arbitrary configurations of the \ch{Au25(SCH3)18} and two isomers of the \ch{Au38(SCH3)24}. Gold-thiolate rings and \ch{Au25(SCH3)18} are especially interesting test case as they are not explicitly included into the training data, hence they demonstrate the generalization possibilities of our ML approach. Furthermore, the existence of gold-thiolate rings in cluster synthesis has been verified experimentally \cite{hostetler98,chen00,corbierre05} and they have also been studied theoretically \cite{gronbeck06}. The tests demonstrate the usefulness of our method for coarse optimization. It can guide optimization to the close vicinity of the local minimum, which can then be reached with finer optimization via DFT. The method allows  breaking and making of chemical bonds, hence in the future it could be applied to the dynamic simulations where chemical reactions can take place.

\section{Computational methods}

Here we go through the theoretical background of the ML approach. First the SOAP descriptor is presented briefly to explain its parameters, which are tested during the model development. Then the background of the distance-based ML methods is introduced and how they are applied to our systems at hand.

\subsection{Smooth Overlap of Atomic Positions}

SOAP is a local descriptor, which means that it is used to describe a local chemical environment of an atom or a single point. The basic idea is to present every atom as a 3D Gaussian function, then present these functions as a series expansion using radial basis functions and spherical harmonics and, finally, collecting coefficient from the expansion into a power spectrum \cite{bartok13, dscribe}. We used the version implemented in DScribe package by Himanen \textit{et al.} \cite{dscribe} and we follow their formalism to introduce main aspects of the SOAP.

The starting point of the SOAP is to represent every atom with a three dimensional Gaussian function. Every element is handled separately and the environment of the point \mbf{r} is written as

\begin{equation}
    \rho^Z (\mathbf{r}) = \sum_i^{\{Z\}} e^{-\frac{|\mathbf{r} - \mathbf{r}_i|^2}{2\sigma_{SOAP}^2}}.
    \label{eq:soap_atom_density}
\end{equation}

\noindent Here $Z$ is an atomic number and the summation goes over all atoms of that type. The positions of these atoms are denoted with $\mathbf{r}_i$. The elegant idea behind SOAP is to use radial basis functions $b_n$ and spherical harmonics $Y_{lm}$ to form a series expansion of the form

\begin{equation}
    \rho^Z (\mathbf{r}) = \sum_{nlm} c^{Z}_{nlm} b_{nl}(r) Y_{lm}(\theta,\phi).
    \label{eq:soap_serie}
\end{equation}

\noindent The coefficients $c^{Z}_{nlm}$ are the heart of the whole description. They are solved via integration

\begin{equation}
    c^{Z}_{nlm} = \int\int\int dV\ b_{nl}(r) Y_{lm}(\theta,\phi)\rho^Z (\mathbf{r})
    \label{eq:soap_coeff}
\end{equation}

\noindent and then collected into a power spectrum

\begin{equation}
    p^{Z_1,Z_2}_{nn'l} = \pi \sqrt{\frac{8}{2l + 1}}\sum_{m} \left(c^{Z_1}_{nlm}\right)^* c^{Z_2}_{n'lm}.
    \label{eq:power_spectrum}
\end{equation}

The values $p^{Z_1,Z_2}_{nn'l}$ are stored into a vector, which works as a local description of the point $\mathbf{r}$. The equation \eqref{eq:power_spectrum} is slightly different than the one in the original publication of Bart\'{o}k \textit{et al.} \cite{bartok13}. In the DScribe package Himanen \textit{et al.} use real (tesseral) spherical harmonics instead of complex ones and, in addition to this, they replace polynomial radial basis functions with Gaussian type orbitals 
 
 \begin{equation}
     b_{nl}(r) = \sum^{n_{\text{max}}}_{n^{'} = 1} \beta_{nn^{'}l}\ r^{l}\ e^{\alpha_{n^{'} l}r^{2} } .
     \label{eq:soap_gaus-orb}
 \end{equation}
 
\noindent This simplifies the theory and makes programming the descriptor efficient. In practise, the summation in the series does not include all indices $n$ and $l$ but they are restricted to maximum values $n_{max}$ and $l_{max}$, which are parameters of the descriptor. The index $l$ restricts the values integer $m$, because $m\in[-l,l]$ same way as side quantum number restrict magnetic quantum numbers. Furthermore, only atoms within some pre-defined cut-off radius $r_{cut}$, which also is a parameter, are included in to the summation in \ref{eq:soap_atom_density}. For further details, see references \cite{bartok13,dscribe}. In this study, we tested the effects of four SOAP parameters: $n_{max}$, $l_{max}$, $r_{cut}$ and Gaussian broadening $\sigma_{SOAP}$.

\subsection{Distance-based ML tools}
\label{sec:dbml}

The basic construct in the distance-based machine learning is to use Euclidean distances between reference and input data as a measure of similarity and to predict an output using these distances. There are two main distance-based ML methods: Minimal Learning Machine (MLM) \cite{desouza15} and Extreme Minimal Learning Machine (EMLM) \cite{karkkainen19}. Both of them are general ML tools and they have been used successfully to predict potential energies for \ch{Au38(SCH3)24} nanoclusters \cite{pihlajamaki20}. Distance-based methods are especially appealing methods to study complex nanostructures, because they have been shown to work well with high-dimensional data and even out-perform deep neural networks in some cases \cite{linja20}. This is due to the distance matrix, which effectively hides the dimensionality of the data. The same feature also makes distance-based ML methods resistant to overfitting \cite{hamalainen19}. In addition to this, distance-based ML methods usually have only one hyperparameter: the number of reference points, which reduces parameter testing. When applying ML methods to nanosystems, there are often several parameters to tune, such as the ones of the descriptors. This means that the user have to optimize the way how the data is presented prior the actual model can be trained. The lack of hyperparameters reduces the need for complex model fitting with different parametrizations of the descriptor.

Recently, a variation of MLM, which specifically addresses the directions of the atomic forces, was proposed: Orientation Adaptive Minimal Learning Machine (OAMLM) \cite{pihlajamaki21}. It takes the concept of using Euclidean distances as an input space similarity measure to perform predictions but instead of predicting corresponding distances to output space references, as MLM does \cite{desouza15}, it predicts cosines of angles between reference vectors and a target vector. It can also produce estimates for the uncertainty of the predictions to provide interesting opportunities for different applications, where uncertainty might play a role. 

We go through the theory behind the distance-based ML methods to form a basis for the discussion on OAMLM and the full force prediction framework. All of these methods start with the input data $\mathbf{X}=\{\mathbf{x}_i\}_{i=1}^{N} \in \mathbb{R}^{N\times n_x}$ and corresponding output data $\mathbf{Y}=\{\mathbf{y}_i\}_{i=1}^{N} \in \mathbb{R}^{N\times n_y}$. In our case, \mbf{X} contains SOAP descriptions of the chemical environments of the atoms and \mbf{Y} information about the forces, either norms or unit vectors pointing to the directions of the force vectors. Let's first consider the simplest method EMLM, which is used to predict the norms of the forces given in \mbf{Y}. From the input data \mbf{X}, $K$ reference points are sampled forming a reference set $\mathbf{Q}=\{\mathbf{q}_j\}_{j=1}^{K} \in \mathbb{R}^{K\times n_x}$.  The training of EMLM is done via regularized least-squares optimization problem, which is used to find optimal weights to perform regression from Euclidean distances between points in $\mathbf{X}$ and $\mathbf{Q}$ to predict $\mathbf{Y}$ \cite{karkkainen19}.

\begin{equation}
    \underset{\mathbf{W}\in \mathbb{R}^{K\times n_y}}{\text{min}} J(\mathbf{W}) = \frac{1}{2N}\sum_{i=1}^{N} \left| \mathbf{d}_i^T\mathbf{W} - \mathbf{y}_i^T\right|^2 + \frac{\beta}{2K}\sum_{i=1}^{K}\sum_{j=1}^{n_y}|W_{ij}|^2.
    \label{eq:rlsq}
\end{equation}

\noindent Vector $\mathbf{d}_i \in \mathbb{R}^{K}$ contains Euclidean distances between $i$th input data point and $K$ references. $\mathbf{W} \in \mathbb{R}^{K\times n_y}$ is a weight matrix, which does a linear regression from kernel space to output. Constant $\beta$ is used for regularization, which might be useful if one has noisy data. In our case, it is fixed to the square root of machine epsilon.

The minimum of the equation \ref{eq:rlsq} can be found by writing it with full matrices and finding the zero point of the first derivative.

\begin{equation}
    \frac{1}{N} \mathbf{D}^T \left( \mathbf{DW} - \mathbf{Y} \right) + \frac{\beta}{K}\mathbf{W} = 0
\end{equation}
\begin{equation}
    \left( \mathbf{D}^T\mathbf{D} +  \frac{\beta}{K}\mathbf{I} \right)\mathbf{W} = \mathbf{D}^T\mathbf{Y}
    \label{eq:emlm_sol}
\end{equation}

\noindent Matrix $\mathbf{D} \in \mathbb{R}^{N\times K}$ contains all Euclidean distances between training data and references. The equation \eqref{eq:emlm_sol} is now a simple representation of the training of EMLM and it can be easily solved numerically. To predict output for an arbitrary input, one has to calculate distances between the input and references forming $\mathbf{d} \in \mathbb{R}^{K}$ and then compute matrix multiplication $\mathbf{d}^T \mathbf{W}$. This is analogous to Kernelized Ridge Regression (KRR), where one has a variety of choices for kernel functions \cite{murphy12}.

Next we shall go through the framework of the MLM presented by de Souza \textit{et al.} \cite{desouza15} and proceed step by step to the direction prediction scheme of the OAMLM. The main difference between MLM and EMLM is that in addition to references $\mathbf{Q}$ in input space MLM also has references $\mathbf{T}=\{\mathbf{t}_j\}_{j=1}^{K} \in \mathbb{R}^{K\times n_y}$ in output space. The idea is not to predict directly output for certain input but to form regression between the two distance spaces.

\begin{equation}
    \mathbf{D}_{out} = \mathbf{D}_{in}\mathbf{B} + \mathbf{\epsilon}.
    \label{eq:mlm-regression}
\end{equation}

\noindent Here $\mathbf{D}_{in} \in \mathbb{R}^{N\times K}$ contains Euclidean distances between the $N$ input training data points in $\mathbf{X}$ and $K$ reference points in $\mathbf{Q}$. $\mathbf{D}_{out} \in \mathbb{R}^{N\times K}$, on the other hand, consists of distances between training output data in $\mathbf{Y}$ and the output reference set $\mathbf{T}$. $\mathbf{B} \in \mathbb{R}^{K\times K}$ is a weight matrix that performs the linear regression and $\mathbf{\epsilon}$ is a residual, which is assumed to be small. It is shown that the approximate solution for the weight matrix is \cite{desouza15}

\begin{equation}
    \mathbf{B} = \left(  \mathbf{D}_{in}^T\mathbf{D}_{in}\right)^{-1}\mathbf{D}_{in}^T\mathbf{D}_{out}.
    \label{eq:mlm-train}
\end{equation}

In order to calculate output with MLM, one first predicts distances between still unknown result and output space references using input space distances and just solved weights as $\mathbf{d}_{out}^T = \mathbf{d}_{in}^T \mathbf{B}$. The result is found by solving multilateration problem, for which there are several methods \cite{navidi98,hamalainen19}.

With these derivations at our disposal, let us proceed to the OAMLM. To remind, the input space training data $\mathbf{X}=\{\mathbf{x}_i\}_{i=1}^{N} \in \mathbb{R}^{N\times n_x}$ contains SOAP descriptions of chemical environments, which do not include directional information. For this reason OAMLM also needs coordinates of neighboring atoms $\mathbf{P}=\{\mathbf{p}_i\}_{i=1}^{N} \in \mathbb{R}^{N\times (1+M)\times 3}$ as an accompanying data. In $\mathbf{p}_i$ the first row is the position of the studied atom itself followed by $M$ neighbors. For every training data point there are also their unit force vectors collected into $\mathbf{Y}=\{\mathbf{y}_i\}_{i=1}^{N} \in \mathbb{R}^{N\times 3}$, where $|\mathbf{y}_i| = 1$ for all values of $i$. Similar to MLM, OAMLM also uses references both in input and output spaces. The $K$ reference data points used are sampled into $\mathbf{Q}=\{\mathbf{q}_j\}_{j=1}^{K} \in \mathbb{R}^{K\times n_x}$ for chemical descriptors, $\mathbf{S}=\{\mathbf{s}_j\}_{j=1}^{K} \in \mathbb{R}^{K\times (1+M)\times 3}$ for coordinates of the neighboring atoms and $\mathbf{T}=\{\mathbf{t}_j\}_{j=1}^{K} \in \mathbb{R}^{K\times 3}$ for corresponding unit force vectors.

Atomic environments can be in any spatial orientation, therefore the directions of the forces cannot be compared directly. As a solution, the coordinates of neighboring atoms in \mbf{P} and \mbf{S} are used to align atomic environments. In this study we used Singular Value Decomposition (SVD) based method presented originally by Arun \textit{et al.} \cite{arun87}. First the atoms, for which forces are predicted, are moved to the origin and their neighbors are translated together with them to preserve the general positioning. Then matrix \mbf{A}$_{i,j}\in \mathbb{R}^{3\times 3}$ is formed by calculating it as

\begin{equation}
    \mathbf{A}_{i,j} = \sum_{k=1}^{1+M} (\mathbf{p}_{i,k}- \mathbf{p}_{i,1}) (\mathbf{s}_{j,k}- \mathbf{s}_{j,1})^{T}.
    \label{eq:svdmat}
\end{equation}

\noindent Index $i$ refers to the $i$th input and $j$ stands for the $j$th reference. With SVD one can split this matrix as $\mathbf{A}_{i,j} = \mathbf{U}\Delta\mathbf{V}^T$. These can be further used to get a rotation matrix $\mathbf{R}_{i,j} = \mathbf{V}\mathbf{U}^T$, which will align points in \mbf{S} and \mbf{P} as well as possible, when the atoms are moved to the origin as in equation \eqref{eq:svdmat}. It is important to notice that this alignment approach depends on the order of given neighborhood points, therefore one needs to form certain rules how the environments are aligned or go through all permutations. However, OAMLM is not restricted to the alignment approach used here. In principle, it is possible to define any alignment scheme suited for specific problems.

The rotation matrices are used to align atomic neighborhoods together, which then yields estimates of alignment accuracy as 

\begin{equation}
    g_{i,j} = \frac{1}{1+M}\sum_{k=1}^{1+M} |(\mathbf{p}_{i,k}- \mathbf{p}_{i,1}) - \mathbf{R}_{i,j}(\mathbf{s}_{j,k}- \mathbf{s}_{j,1})|
    \label{eq:success1}
\end{equation}

\noindent or 

\begin{equation}
    g_{i,j}^{\prime} = \frac{1}{1+M}\sqrt{\sum_{k=1}^{1+M} |(\mathbf{p}_{i,k}- \mathbf{p}_{i,1}) - \mathbf{R}_{i,j}(\mathbf{s}_{j,k}- \mathbf{s}_{j,1})|^2}.
    \label{eq:success2}
\end{equation}

\noindent The same rotation matrices are also used to rotate reference unit force vectors in \mbf{T} to be comparable with data in \mbf{Y}. Dot products between these vectors are calculated as $\hat{\mathbf{y}}_i\cdot (\mathbf{R}_{i,j}\hat{\mathbf{t}}_j)$. This dot product is the cosine of the angle between two vectors, as we are working with unit vectors, and it is evaluated by OAMLM during the prediction phase \cite{pihlajamaki21}. The $g^{(\prime)}_{i,j}$ and dot products are used to form matrices $\mathbf{D}_{g} = \{g^{(\prime)}_{i,j}\} \in \mathbb{R}^{N\times K}$ and $\mathbf{D}_{c} = \{\hat{\mathbf{y}}_i\cdot (\mathbf{R}_{i,j}\hat{\mathbf{t}}_j)\} \in \mathbb{R}^{N\times K}$ respectively.

Now one has everything needed to train the OAMLM using the same training scheme as for MLM in equation \eqref{eq:mlm-train}. \mbf{D}$_{in}$ is the same as before: Euclidean distances between datapoints in \mbf{X} and \mbf{Q}. However, \mbf{D}$_{out}$ is different. As mentioned in the reference \cite{pihlajamaki21}, OAMLM has two weight matrices: \mbf{B}$_{c}$ to predict dot products and \mbf{B}$_{g}$ to predict alignment successes. To acquire those \mbf{D}$_{out}$ in equation \eqref{eq:mlm-train} is substituted with \mbf{D}$_{c}$ or \mbf{D}$_{g}$ correspondingly. However, in this study we do not use \mbf{B}$_{g}$, which could be used for uncertainty estimation. We use only \mbf{B}$_{c}$ to predict dot products.

\begin{figure}
    \centering
    \includegraphics[width=\columnwidth]{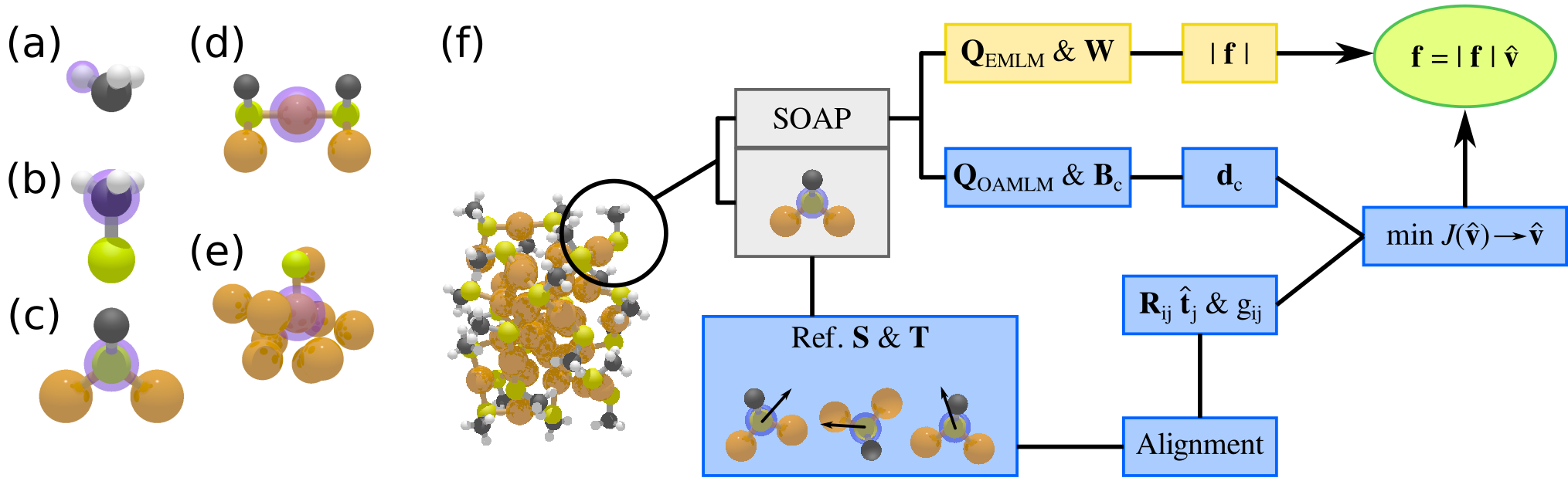}
    \caption{The atomic force prediction framework. Examples of atomic environments used in alignment for (a) hydrogen, (b) carbon, (c) sulfur, (d) unit gold and (e) core gold. The atoms, for which the alignment is done, are highlighted with purple. Panel (f) demonstrates the full force prediction scheme. Description part is shown in grey boxes, norm prediction with EMLM in yellow and the direction estimation of the OAMLM in blue boxes. Colors for atoms: orange, gold; yellow, sulfur; gray, carbon; white, hydrogen.}
    \label{fig:fullmodel}
\end{figure}

The output prediction procedure with OAMLM is similar to the methods in MLM. The schematic picture of the full force prediction process is shown in the FIG. \ref{fig:fullmodel}. As an input, the method takes description \mbf{x}$_i$ and its neighborhood coordinates \mbf{p}$_i$. Vector \mbf{d}$_{in}$ is formed by calculating Euclidean distances between \mbf{x} and the reference points in \mbf{Q}. The weight matrix $\mathbf{B}_{c}$ is used to predict dot products as $\mathbf{d}_{c}^T = \mathbf{d}_{in}^T \mathbf{B}_{c}$. Then reference neighborhood coordinates in \mbf{S} are are aligned with \mbf{p} yielding alignment accuracies $g^{(\prime)}_{i,j}$ and with corresponding rotation matrices reference unit vectors in \mbf{T} are rotated accordingly. The last part of the prediction is similar to the multilateration problem. However, instead of minimizing the distance differences we minimize the difference between the predicted dot products and dot products of the rotated reference unit force vectors $\mathbf{R}_{i,j} \hat{\mathbf{t}}_{j}$ and yet unknown vector $\hat{\mathbf{v}}$. There is no specific method to do this. In the reference \cite{pihlajamaki21} the $\hat{\mathbf{v}}$ was found numerically by using Sequential Quadratic Programming (SQP) to optimize cost function

\begin{equation}
    \underset{\hat{\mathbf{v}}_i\in \mathbb{R}^{3}}{\text{min}}\ J_{1}(\hat{\mathbf{v}}_i) =
    -\sum_{j=1}^K \text{exp}\left(-\left(\frac{d_{c,j} - (\mathbf{R}_{i,j}\hat{\mathbf{t}}_j )\cdot \hat{\mathbf{v}}_i}{\sigma_1} \right)^2 - \left(\frac{g^{(\prime)}_{i,j}}{\sigma_2} \right)^2\right),
    \label{eq:numloss}
\end{equation}

\noindent We call this a numeric loss function. Here we do not make initial selection of the used reference data as in the original paper \cite{pihlajamaki21} but we simply use all references.

In this study we decided to also use more simple cost function as a comparison:

\begin{equation}
    \underset{\hat{\mathbf{v}}_i\in \mathbb{R}^{3}}{\text{min}}\ J_{2}(\hat{\mathbf{v}}_i) =    \frac{1}{2}\sum_{j=1}^{K} \omega_{i,j}\left[ \hat{\mathbf{v}}_i\cdot(\mathbf{R}_{i,j}\hat{\mathbf{t}}_j ) - d_{c, j}\right]^2 ,
    \label{eq:anloss}
\end{equation}

\noindent where

\begin{equation}
    \omega_{i,j} = \text{exp}\left(-\left(\frac{g^{(\prime)}_{i,j}}{\sigma_2} \right)^2\right).
    \label{eq:loss_weighting}
\end{equation}

\noindent The advantage of equation \eqref{eq:anloss} is that it can be solved analytically by taking a derivative over $\hat{\mathbf{v}}_i$ and as a result

\begin{equation}
    \hat{\mathbf{v}}_i = \frac{\sum_{j=1}^{K}\omega_{i,j} d_{c, j} (\mathbf{R}_{i,j}\hat{\mathbf{t}}_j )}{\sum_{j=1}^{K}\omega_{i,j}}.
    \label{eq:analytic}
\end{equation}

\noindent The result is interestingly a weighted average of predicted projections. In practise, $\hat{\mathbf{v}}_i$ is not a unit vector, because there is always numeric error present in the values of \mbf{d}$_{c,j}$ and $\omega_{i,j}$, therefore one has to remember to divide it with its norm before using the result.  We call equation \eqref{eq:anloss} as an analytic loss function. In these two loss functions, $\sigma_1$ and $\sigma_2$ are parameters of the ML model and they are also tested during the model development.

\subsection{Atomic force prediction scheme for \ch{Au38(SCH3)24}}
\label{sec:au38scheme}

\ch{Au38(SCH3)24} nanocluster, which is shown in FIG. \ref{fig:orig_struc}, contains four different elements and has chemically various environments. There are covalently bound methyl thiolate ligands. There is a metallic gold core, where gold atoms are interacting with each other. On the surface of the core some gold atoms can also form bonds to the sulfur atoms. Within the metal-ligand interface structure, sulfur and gold atoms are bound with relatively covalent nature forming protecting units. Within these units the gold atoms are bound only to sulfur atoms, ideally forming two Au-S bonds. There are very diverse features determining the interactions between atoms, therefore it is a good idea to split the problem into smaller parts.

We classify the atoms into five categories: core gold atoms inside the metallic core, unit gold atoms in protecting units, sulfur, carbon and hydrogen. For every atom type we train one EMLM for force norms and one OAMLM for force directions. The norm prediction part is a straightforward standard ML problem, where the method predicts a scalar output according to a given input and the references. For the direction scheme, we have to define, which neighborhood atoms are used to align reference environments to an input environment. In principle, one could just select $n$ nearest neighbors and go through all permutations. However, this wastes computational resources by attempting many unfavorable permutations. Hence, we need to define certain rules according to physical and chemical understanding.

The most simple alignment scheme is for hydrogen. It uses only the nearest carbon, and two other nearest hydrogen atoms bound to the carbon as seen in the FIG. \ref{fig:fullmodel} (a). There are only two permutations of the hydrogen atoms to test. Aligning carbon is similar to the hydrogen scheme. It uses the nearest sulfur atom and three hydrogen atoms, as shown in the FIG. \ref{fig:fullmodel} (b), which results into six permutations of hydrogen atoms to be tested. The alignment of a sulfur atom neighborhood uses the nearest carbon and two nearest gold atoms shown in the FIG. \ref{fig:fullmodel} (c). There are only two permutations of the gold atoms to be tested. 

The gold atoms have the most versatile chemical environments of all atoms in the cluster. Unit gold atoms use two blocks of atoms for alignment. The blocks contain the nearest sulfur atom and two other atoms bound to it: a carbon and another gold atom. Hence, there are two sulfur, two carbon and two gold atoms used to do the alignment. An example of the neighborhood structure is visualized in FIG. \ref{fig:fullmodel} (d). These atoms are handled as blocks, due to the linear nature of the S-Au-S bonding, therefore there are only two permutations to test. 

The MD data used in model development is extremely dynamic and the nature of the Au-S bonds might change significantly. Hence, if a gold atom has only one sulfur within $3.0$ Å and there is no another gold atom within the same distance, the gold atom is considered to be just a half of an unit. This corresponds to a transition state where old unit is broken and new is going to be formed. In this case alignment is done by using only one block of sulfur, carbon and gold atoms. This kind of alignment is much more unstable than the standard way but fortunately breaking of S-Au bond is not a common phenomenon. For hydrogen, carbon, sulfur and unit gold atoms the alignment accuracy is calculated using the equation \eqref{eq:success1}.

The environments in the metallic core gold atoms can be very homogeneous making alignment difficult. Within the core the gold atoms can be bound to a single sulfur atom and the rest of the interactions are metallic or another scenario is that all interactions are metallic. For every core gold there can be maximum of twelve neighboring atoms selected. If there is a sulfur atom within $3.0$ Å, it will be selected first. Then the rest are nearest gold atoms within $5.0$ Å from the nearest to the furthest. There can be less than twelve neighbors selected for a core gold atom, if there are not so many fulfilling the requirements as seen in the FIG. \ref{fig:fullmodel} (e). It is clear that there are too many neighboring atoms to go through all possible permutations in a reasonable amount of time. There can be maximum $12! = 479 001 600$ permutations for a single atomic neighborhood. In order to make the task feasible, we device two alignment schemes depending on whether the aligned gold atoms are bound to a sulfur atom or not. Alignment schemes use always the coordinates of the main gold atom itself and two neighbors, which reduces alignment time significantly. The goal is not to make perfect alignment but make as systematic as possible.

The first scenario for core gold is that both input and reference gold atoms have a sulfur atoms within their immediate vicinity. In this case, the alignment is done by using three points: gold atom itself, sulfur atom and one neighboring gold atom. For input environment we select the nearest neighboring gold atom as the third point. For reference environment the selection is the same except that in addition to the nearest neighboring gold atom we also go through all other possible neighboring gold atoms. These three atoms are used to make alignments and the accuracy is evaluated with equation \eqref{eq:success2}.

The second scenario is that at least one of the environments does not contain sulfur. Here the alignment uses only three atoms similarly to the previous core gold scenario. For the input environment the three points are the atom itself and its two nearest neighbor gold atoms. For the reference environment, we use the atom itself and all possible pairs of the neighbors. Here the order does play a role, therefore one would get maximum of $12\cdot 11 = 132$ pairs.

These pairs together with the main gold atom form triangles, which are used to rule out some permutations. The reasoning for this ruling out is to reduce computation time. Comparing triangles is faster than making actual SVD alignments and measuring the accuracy. The difference between the $k$th triangle of the reference environment and the triangle formed from input data is measured as

\begin{equation}
    u_k = \sum_{i=1}^{3}[(l_{k, i} - l_{0, i})^2 + (\theta_{k, i} - \theta_{0, i})^2].
    \label{eq:triangle_comparison}
\end{equation}

\noindent Here $l_{k, i}$ is the length of the $i$th side of the triangle in \r{A}ngstroms and $\theta_{k,i}$ is an angle of the $i$th corner in radians. The lower index $k$ refers to the reference data triangle and lower index $0$ to the input data triangle. Then $n$ triangles, for which the difference $u_k$ is the smallest, are selected. We decided to use $n=10$. These triangles are used to make SVD alignments and the one yielding the smallest value of the equation \eqref{eq:success2} is selected. These two alignment schemes for core gold atoms are designed to be as similar as possible, because they will be used by the same OAMLM. If the alignment would vary much between different core gold atoms, it would make the direction estimation even more difficult for the method than it is currently.

As mentioned earlier, the number of neighborhood atoms for the core gold atoms is not constant. Hence, when the alignment success is estimated, it is required that every atom has some nearest neighbor distance. Let us clarify this via an example. If input environment has 6 neighbors and reference has 10, then after the alignment we measure the nearest neighbor distance for all 10 atoms in the reference environment and use them in the equation \eqref{eq:success2}. It does not matter whether reference or input has more atoms but the accuracy is always estimated with the largest number of nearest neighbor distances. This is used to emphasize the differences between the atomic environments of the core gold atoms.

Implementing chemical rules and primary knowledge into the algorithm resembles the approach to construct metal-ligand interfaces by Malola \textit{et al.} \cite{malola19}. There authors used distances and angles to compare environments between reference structures and the environments of arbitrary points within unprotected metal clusters. This comparison enabled them to determine whether or not those points were suitable for interface atoms. Our force prediction method shows similar philosophy to the task but here we have to use actual spatial alignment in order to capture the orientation information.

\subsection{Structure optimization via ML forces}

As an usage example of ML forces, we perform structure optimization in different scenarios. The model does not yield values for potential energy of the system but the optimization is run solely with ML estimated forces. We used classic quasi-Newton method Broyden–Fletcher–Goldfarb–Shanno (BFGS) algorithm \cite{broyden70,fletcher70,goldfarb70,shanno70} to run structure optimization. The challenge is that ML predicted forces have always some level of uncertainty, which is seen in the optimization algorithm as a noise. In this study we do not explicitly address the uncertainty in the optimization algorithm but it is an aspect that should be considered in the future studies. The used BFGS implementation is based on the one included in Atomic Simulation Environment (ASE) package \cite{ase17}.

\subsection{DFT methods}

For reference calculations, we used the DFT code GPAW \cite{gpaw} as it was also used in the original MD simulations of \ch{Au38(SCH3)24} by Juarez-Mosqueda \textit{et al.} \cite{rosalba19}. The exchange-correlation functional was Perdew-Burke-Ernzerhof functional (PBE) \cite{pbe} and we used $0.2\ \text{\r{A}}$ real space grid spacing. BFGS structure optimization using GPAW computed potential energies and forces were run with the original implementation in ASE package \cite{ase17}. The DFT-level BFGS optimizations were considered to be converged if the maximum force of the atoms was $\leq0.05\ \text{eV/Å}$.

\section{Results and discussion}

The results are divided into six parts. First the effect of SOAP parameters to norm and direction prediction are shown. This way the optimal description parameters are found. They are used in the next two parts, where full EMLM models for norms and OAMLM models for directions are trained and tested. The last three parts focus on structure optimization. The used test cases are gold-thiolate rings, \ch{Au38(SCH3)24} cluster structures with outstretched protecting units and snapshots from the MD simulations of the \ch{Au25(SCH3)18} and of the both isomers of the \ch{Au38(SCH3)24} nanocluster.

The training and testing of the models relies heavily on the DFT-level MD simulation data of the \ch{Au38(SCH3)24} nanocluster from reference \cite{rosalba19}. In that study, authors run long MD simulations on both isomers of the \ch{Au38(SCH3)24}, where the systems were heated from $0\ \text{K}$ to $1100\ \text{K}$ so that they broke down. The less stable T isomer started to undergo significant structural changes very early and in the later stages highly deformed seven gold atom gold-thiolate ring broke out of the structure. These simulations resulted into over 12\ 000 configurations for both isomers, which serve as an ideal dataset for our study here.

\subsection{Data and SOAP parameter selection}

The data used to train and test our model was extracted from the DFT level MD simulations of \ch{Au38(SCH3)24} published in the reference \cite{rosalba19}. For both isomers we sampled 1000 configurations logarithmically, which means that configurations were sampled sparsely from the beginning of the trajectory and more densely from the end in a similar fashion as log-scale graphs are plotted. This guaranteed that we got denser sampling from the high temperature region, where there are more changes in the structures, than from the low temperature region. This data contains 24 000 local environments for carbon and sulfur, 72 000 for hydrogen from both isomers. Q isomer data contains 22 836 core, 15 123 unit and 41 half unit gold atoms. T isomer data contains 22 055 core, 15 888 unit and 57 half unit gold atoms. Unit and half unit gold atoms are handled by the same EMLM and OAMLM models.

The importance of the level of description cannot be emphasized too much. If description is not accurate enough the prediction will be poor. However, if description is overly accurate, it will lead to a highly specialized model, which cannot be generalized and the risk of overfitting increases. We tested several SOAP parameters: $r_{cut}\in \{4.0 \text{Å},\ 5.0 \text{Å}\}$, $\sigma_{SOAP}\in\{1.0, 0.75, 0.5, 0.25\}$, $n_{max}\in[2,7]$ and $l_{max}\in[0,4]$. This totals $240$ description sets for sulfur, carbon and hydrogen. For gold atoms we used only $\sigma_{SOAP} = 0.25$ sets resulting $60$ SOAP parameter sets. In this article and its Supplemental Material, we show only a selected portion of the tests. The complete analysis of the parameter tests is available in (\url{https://gitlab.jyu.fi/aneepihl/oamlm_forces.git}).

First these sets were used to predict norms of the forces and to restrict the number of parameters to be tested in the  direction prediction scheme. It is easier to predict norms than directions, therefore it is reasonable to assume that if norms are predicted inaccurately directions won't be any better. For every parameter set we trained one EMLM with Q isomer data and one EMLM with T isomer data. Then we used Q model to predict norms from T set and vice versa. This is close to so-called cross validation approach often used when testing ML methods. 

From every data set of a single isomer, 2500 points were selected with RS-maximin sampling \cite{gonzalez85,hamalainen19}. Here data points refer to the local atomic environments and their corresponding forces. This data was used as a training data and all points were saved as references into the models. The SOAP data points were minmax scaled between 0 and 1. The performance was measured with root mean squared error (RMSE). Tests showed the most promising parameters to be $\sigma_{SOAP}=0.25$, and $(n_{max},l_{max}) \in\{(6,4), (7,3), (7,4)\}$ with both $r_{cut}=4.0\ \text{Å}$ and $r_{cut}=5.0\ \text{Å}$ resulting to only six parameter sets to be tested with OAMLM. The results with $\sigma_{SOAP}=0.25$ and $r_{cut}=4.0\ \text{Å}$ are shown in the Supplemental Material  figures $\text{S}1-\text{S}4$ for sulfur, $\text{S}5-\text{S}8$ for carbon, $\text{S}9-\text{S}12$ for hydrogen, $\text{S}13-\text{S}16$ for unit gold and $\text{S}17-\text{S}20$ for core gold.

Testing with OAMLM was done in a similar fashion as with EMLM: models were trained with one isomer and then tested with another. As mentioned in the section \ref{sec:dbml}, the output direction estimation can be done via numeric or analytic loss function by using either equation \eqref{eq:numloss} or \eqref{eq:anloss}. The initial tests were ran with both output estimation methods and their parameters were set as $\sigma_1=0.25$ and $\sigma_2=0.5$. The performance was measured with weighted average of the angles between the estimated force directions and the corresponding DFT calculated force vectors. The squared norms of the DFT force vectors worked as weights. This emphasizes the handling of the large forces, for which it is more important to get directions correct than for the small ones. When the norm decreases the direction of the force vector becomes more and more elusive and sensitive to changes in the chemical environment.

The analytic loss function was performing better than the numeric one, which showed unstable performance. Parameters $\sigma=0.25$, $n_{max}=7$, $l_{max}=4$ and $r_{cut}=4.0\ \text{Å}$ showed satisfying performance for all atom types. The results with these parameters using numeric loss function are shown in Supplemental Material figure $\text{S}21$ and analytic loss function results are shown in $\text{S}22$. The longer cut-off radius did not show a significant improvement compared to the selected one, therefore it is natural to use shorter one. There will be less atoms included into the description making it slightly faster to compute and it is more likely to results in generalizable method.

After finding the optimal SOAP parameters, we also tested how $\sigma_2$ parameter affects the performance of the analytic loss function. In addition to the previously used value of $0.5$, we also tested values $0.25$ and $0.75$ with previously acquired SOAP parameters. The results with these parameters are shown in Supplemental Material figures $\text{S}23$ and $\text{S}24$. The tests do not show any significant effect to better or worse. For unit gold atoms the $\sigma_2 = 0.25$ seem to be slightly better option than $0.5$, because the weighted average angles were previously approximately $29^\circ$ $(\text{Q}\rightarrow\text{T})$ and $25^\circ$ $(\text{T}\rightarrow\text{Q})$, and with smaller $\sigma_2$ parameter the values decreased to about $25^\circ$ $(\text{Q}\rightarrow\text{T})$ and $24^\circ$ $(\text{T}\rightarrow\text{Q})$. We settled on $\sigma_2 = 0.25$ for unit gold atoms and for everything else $\sigma_2 = 0.5$.

\subsection{Full EMLM force norm models}

\begin{figure*}
    \centering
    \includegraphics[width=\columnwidth]{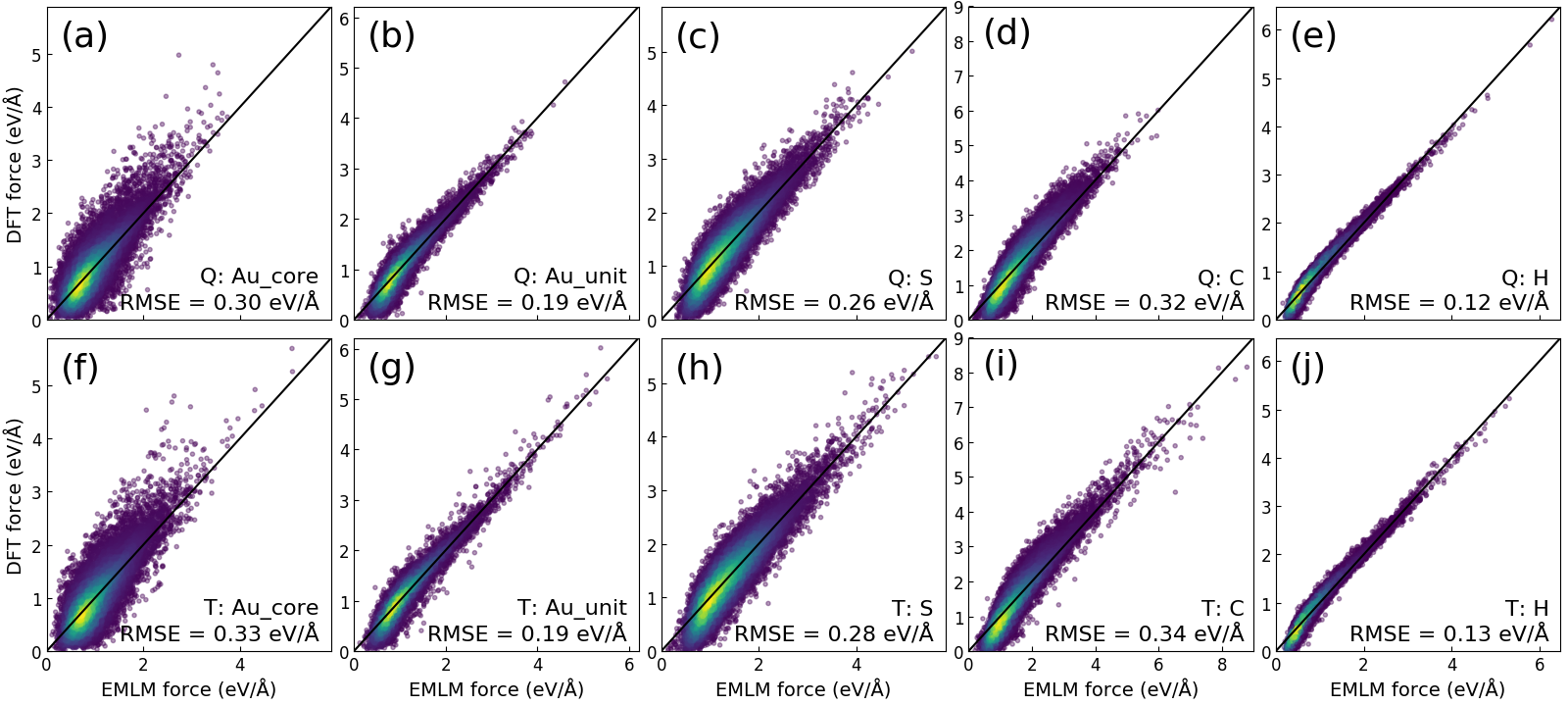}
    \caption{Performance of different full EMLM models in comparison to DFT level forces. Panels (a)-(e) show the test results for the Q isomer and (f)-(j) for the T isomer. The tested element is written to the corner of every graph along with RMSE values. For hydrogen only third of the data points are plotted. The colors visualize the density of the points: yellow means dense region and purple sparse.}
    \label{fig:emlm_norms}
\end{figure*}

After determining suitable parameters for the SOAP description, we trained EMLM with combination of data from both Q and T isomers. From the combined data set, 5000 points for each atom type were selected with RS-maximin sampling \cite{gonzalez85,hamalainen19}. This data was used as a training data and all points were saved as references. All descriptions were minmax scaled between 0 and 1. The models were tested with the remaining data from both isomers. The predictions are visualized in FIG. \ref{fig:emlm_norms} along with RMSE values.

For unit gold, sulfur and hydrogen RMSEs are lower than $0.3 \text{eV/\r{A}}$ and predictions correlate well with DFT force norms as seen in FIG. \ref{fig:emlm_norms} (b), (c), (e), (g), (h) and (j). The largest RMSE values belong to core gold and methyl carbon model. It is expected that gold core is difficult to handle as it undergoes great deal of structural changes. Against expectations, methyl carbon proved to be difficult for the EMLM. The chemical environment of the carbon is mostly determined by its neighboring hydrogen atoms and a sulfur, therefore the changes are quite small due to the rigid covalent bonds. A logical explanation would be that the carbon needs more exact SOAP description with high sensitivity to small changes. This could be achieved by using even smaller value for Gaussian broadening parameter $\sigma_{SOAP}$. However, the acquired accuracy is reasonable and can be used as a part of the simulations.

\subsection{Full OAMLM force direction models}

The OAMLM models were trained in same manner as EMLM but only 2500 data points were used in training and as references. The alignment of atomic environments is a relatively slow process, therefore having fewer references makes the model more feasible to use. During the predictions the weighting parameter in analytic loss function \eqref{eq:anloss} was set as $\sigma_2=0.25$ for unit gold atoms and for everything else $\sigma_2=0.5$. As an error measurement we used weighted average of angles between predicted force directions and DFT level force vectors. As a weights, we used the squared norms of the DFT forces the same way as before. 

\begin{figure*}
    \centering
    \includegraphics[width=\columnwidth]{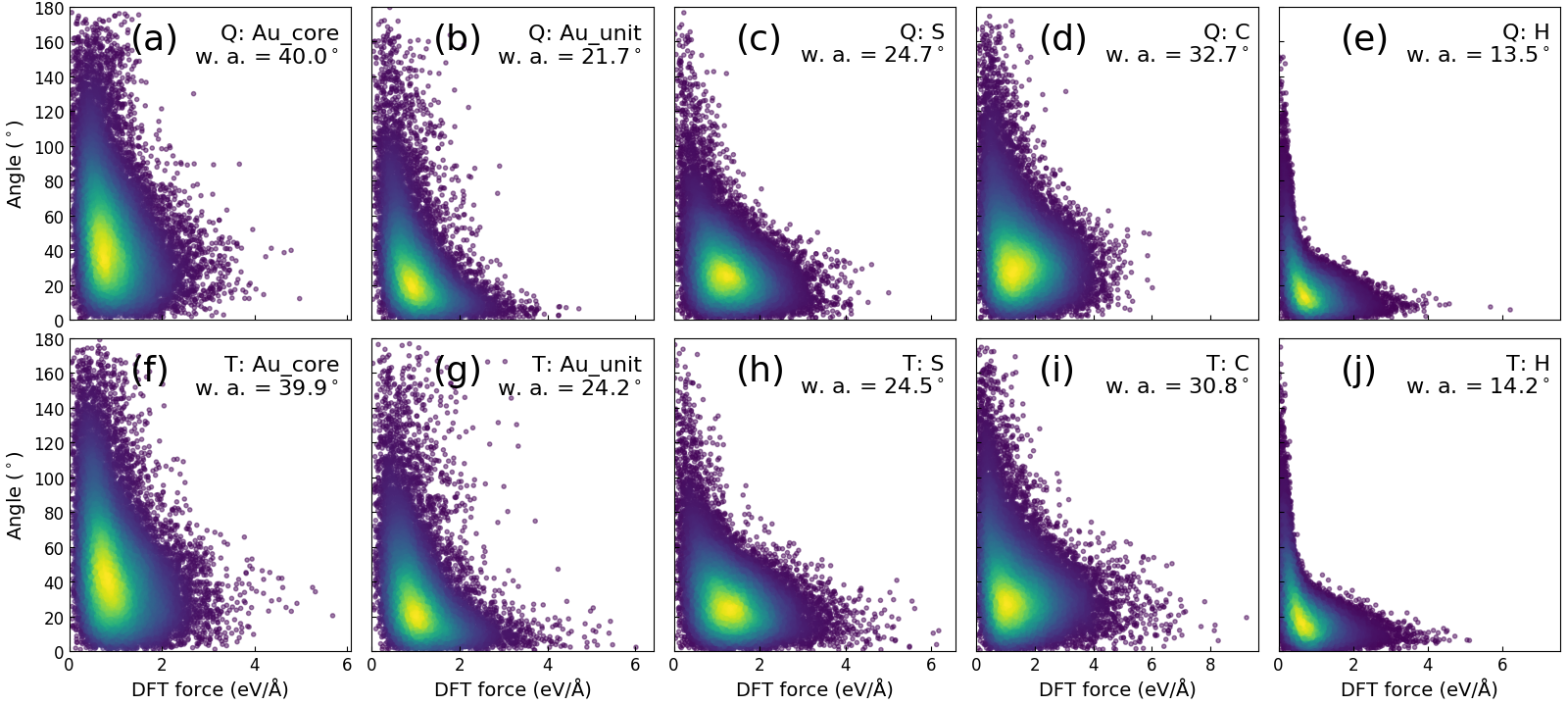}
    \caption{Performance of the full OAMLM models using analytic loss function in equation \eqref{eq:anloss}. Vertical axes are the angle between the predicted direction and the DFT force vectors. Horizontal axes show corresponding DFT force norms. Panels (a)-(e) show the test results for the Q isomer and (f)-(j) for the T isomer. The tested element is written to the corner of every graph. For hydrogen only third of the data points are plotted. In the graphs, "w. a." stands for weighted average. The colors visualize the density of points: yellow means dense region and purple sparse.}
    \label{fig:oamlm_test}
\end{figure*}

The results are plotted in the FIG. \ref{fig:oamlm_test}. The effects of small forces are visible in all plots. When the norm of the force is small, the direction is extremely difficult to be estimated, which leads to the increased deviation close to the zero. The weighted averages show similar trends as the RMSEs in the case of force norms. Unit gold, sulfur and hydrogen are the easiest to handle as seen in FIG. \ref{fig:oamlm_test} (b), (c), (e), (g), (h) and (j). From these three atom types the largest the largest weighted average angle $24.7^\circ$ belongs to sulfur atoms of the isomer Q. The unit gold data contains some individual points, for which the angle is not as accurate as for the rest. This uncertainty is most likely caused by the inclusion of "half unit" gold atoms and possible classification difficulties. The classification rules mentioned in the section \ref{sec:au38scheme} are approximate and especially the T isomer data might contain instances where classification is not clear. 

For core gold atoms in the FIG. \ref{fig:oamlm_test} (a) and (f) the points are more spread than the other atom types. This hints that the alignment of the core environment is not straightforward, which leads into difficult estimation of the direction. However, OAMLM still manages to yield reasonable estimates even with highly complex alignment situations. For the methyl carbons in the FIG. \ref{fig:oamlm_test} (d) and (i), the origin of the uncertainty is likely the same as in the case of force norm prediction. It needs more exact SOAP description with small gaussian broadening parameter $\sigma_{SOAP}$. On the other hand, it is not easy to make unambiguous alignment for the very symmetric methyl group. Hydrogen atoms do not make extreme movements, therefore all their permutations yield very similar alignments, which are difficult to distinguish without highly specialized structural descriptors.

\subsection{Application to structure optimization}

As we now  have a full force estimation method combined from EMLMs and OAMLMs, the next step is to apply it to the structure optimization with BFGS. In the first test we leave the complicated metallic core out and focus on covalently bound parts by optimizing gold-thiolate rings. The second case is to optimize a stretched protecting unit attached to the \ch{Au38(SCH3)24} cluster. The third one is the most difficult test, where we use our model and BFGS to optimize snapshots from the MD simulation trajectories of $[$\ch{Au25(SCH3)18}$]^-$ and \ch{Au38(SCH3)24}.

\subsubsection{Gold-thiolate rings}

\begin{figure}
    \centering
    \includegraphics[width=0.4\columnwidth]{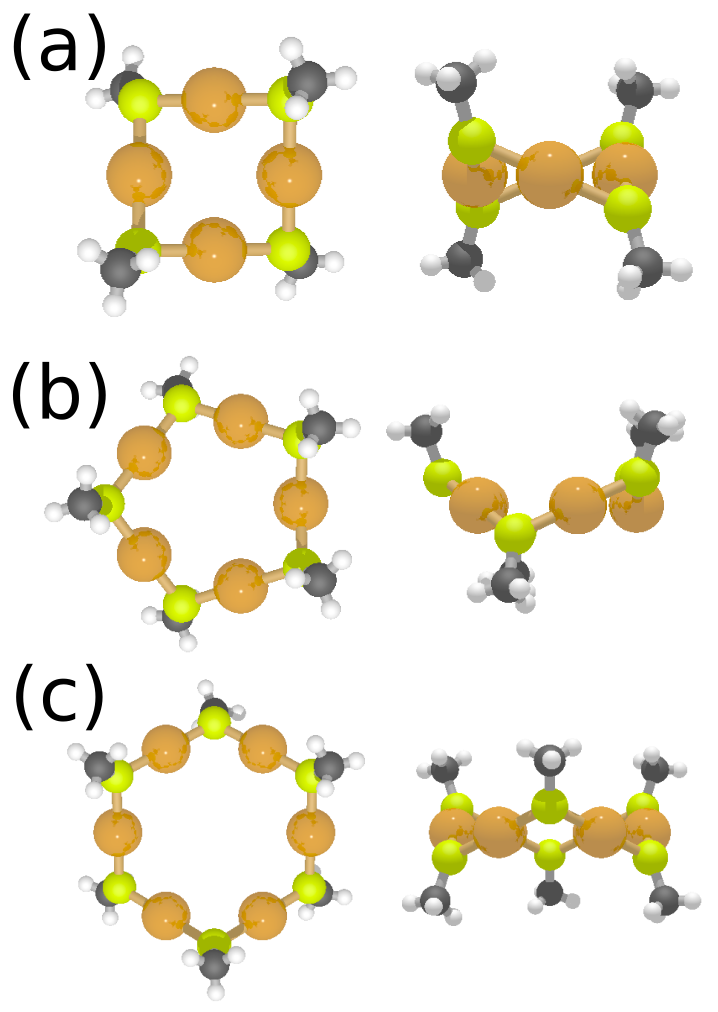}
    \caption{Top and side views of the initial structures for (a) four, (b) five and (c) six gold atom gold-thiolate rings. Colors: orange, gold; yellow, sulfur; gray, carbon; white, hydrogen.}
    \label{fig:initial-rings}
\end{figure}

Testing the model with gold-thiolate rings is an interesting test case, because the model is not explicitly trained with them. In the MD trajectory of the T isomer there is an seven gold atom ring breaking out from the structure in the end \cite{rosalba19} but there is no guarantee how much it has been sampled and the ring in MD is highly deformed. The starting structures were generated by making even geometric shapes, where sulfur atoms lie in the corners. Sulfur atoms were displaced from the plane $1.0 \text{\r{A}}$ up and down in turns. We focus on the rings containing four, five or six gold atoms. These structures are shown in FIG. \ref{fig:initial-rings}.

Structures were optimized by both DFT and ML model using BFGS algorithm. Optimization with DFT used the default $0.2\ \text{\r{A}}$ maximum step size of the ASE package. For ML forces the step size was set to half smaller value of $0.1\ \text{\r{A}}$. ML-based optimization ran 200 optimization step, which was its maximum number of iterations. The stopping criterion was that if maximum force is $\leq 0.1\ \text{eV/\r{A}}$, the optimization would stop. However, due to the uncertainty in the model affecting the behavior of BFGS algorithm, optimization did not reach this limit. 
Even if the optimization is almost converged, BFGS will still propose improving steps. In some cases this might cause individual atoms to be moved slightly away from their local minimum increasing the forces. This is seen as a small fluctuation when the optimization is at the verge of convergence. After optimizations, potential energies were computed for ML optimization trajectories via single point DFT calculations. 

\begin{figure*}
    \centering
    \includegraphics[width=\columnwidth]{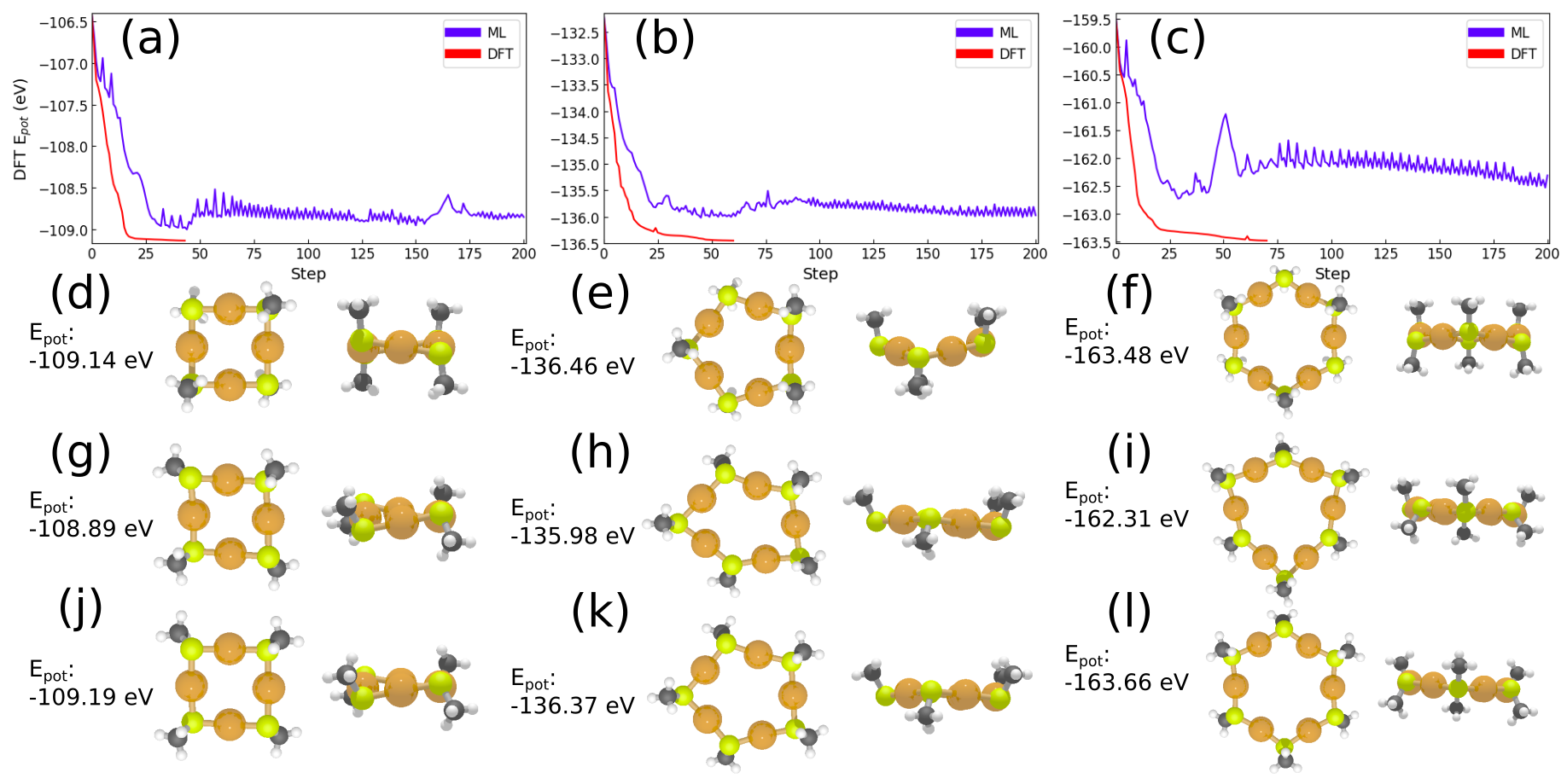}
    \caption{(a)-(c) show the DFT calculated potential energy evolution during the DFT and ML BFGS optimization for four, five and six gold atom gold-thiolate rings respectively. (d)-(f) the final structures from the DFT optimization viewed from top and side. Correspondingly (g)-(i) are the final structures from ML optimization. Structures in (j)-(l) are DFT optimization results, which started from the corresponding ML optimization results. Colors: orange, gold; yellow, sulfur; gray, carbon; white, hydrogen.}
    \label{fig:rings}
\end{figure*}

The potential energies in FIG. \ref{fig:rings} panels (a), (b) and (c) are decreasing during the optimization as supposed to. For four and five gold atom rings, the descending of the potential energy is effectively monotonous. With six gold atoms, the ML optimization initially manages to decrease the potential energy the same manner as before but after about 50 steps it adopts a geometry, which does not fully agree with DFT. The six gold atom ring contains more empty space in the middle of the ring than in the any configuration used to train the ML model, therefore it is expected that increasing the ring size increases the uncertainty of the model.

The comparison of the structures from DFT and ML optimization reveals  intriguing differences. The four gold ring configuration, into which DFT optimization converged, is just slightly twisted clockwise out of the plane as seen in FIG. \ref{fig:rings} (d). However, ML optimization has been twisted on the opposite direction in FIG. \ref{fig:rings} (g). Similar trend is also seen with five gold atom ring in FIG. \ref{fig:rings} panels (e) and (h). For six gold atom ring, the twisting is not very clear in FIG. \ref{fig:rings} panels (f) and (i). There the hexagonal ring shape has been deformed towards the triangle, which is likely caused by the method preference to produce $90^\circ$ Au-S-Au angles locally as ML methods do not see the whole structure.

Due to the differences in the DFT and ML optimization results, we decided to optimize the final structures from the ML optimization with DFT. The results are shown in FIG. \ref{fig:rings} panels (j), (k) and (l). It is surprising that in the case of four and six gold atom rings the potential energies of these newly optimized structures are slightly better than the ones from direct DFT optimization. The twisting has also been preserved, which indicates that the structural differences in plain DFT and ML optimizations are not defects but features of realistic local energy minima. 

The structures optimized only with DFT settled to a local energy minimum close to the initial structures. ML method on the other hand passed this minimum and continued into another one resulting into an opposite twisting of the structure. It is likely that the firstly mentioned energy minimum is shallow compared to its surrounding potential energy landscape. The ML method either has not learned this kind of profile or the minimum was hidden by the uncertainty in the model. However, this behavior enabled the optimization to proceed close to an alternative energy minimum, which could possibly be even better. This demonstrates that our ML methodology can be utilized as a hybrid optimization tool, where ML executes coarse optimization and DFT is used in fine tuning.

\subsubsection{Partial optimization of the \ch{Au38(SCH3)24} nanocluster}

\begin{figure}
    \centering
    \includegraphics[width=0.6\columnwidth]{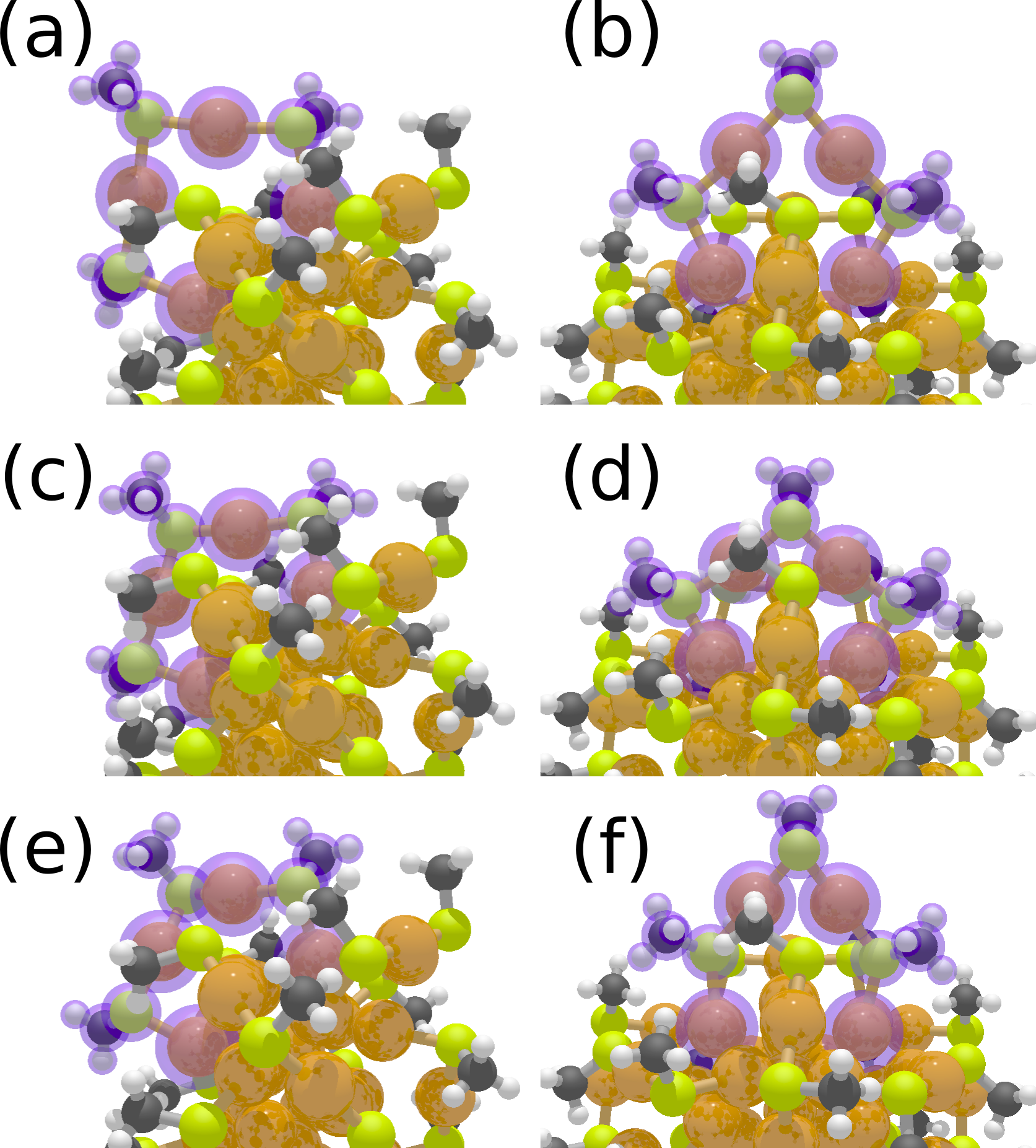}
    \caption{(a) and (b) show the stretched protecting units the \ch{Au38(SCH3)24} Q and T isomer respectively. (c) and (d) are DFT constrained optimization result starting from the structures (a) and (b). (e) and (f) are constrained ML optimized structures from 150th optimization step with $0.05 \text{\r{A}}$ maximum BFGS step size. During the optimization everything else is fixed expect the part highlighted with purple. Colors: orange, gold; yellow, sulfur; gray, carbon; white, hydrogen.}
    \label{fig:au38struc}
\end{figure}

The second test case is to optimize \ch{Au38(SCH3)24} structures, which are otherwise DFT optimized except one long protecting unit is pulled outwards $2.0\ \text{\r{A}}$. This is done for both isomers. As seen in the FIG. \ref{fig:au38struc} (a) for Q isomer the pulled unit lies on the corner of the cylindrical shape and for T isomer the pulled unit is in the middle of the structure presented in \ref{fig:au38struc} (b). As a comparison to the ML optimization, we optimized the structure also with DFT forces and BFGS. During the optimization process only atoms belonging to stretched unit were allowed to move and others were fixed, therefore there were four gold, three sulfur, three carbon and nine hydrogen atoms that are moving. Two of the gold atoms were classified as belonging to the unit and two to the core.

Different maximum step sizes of the ML BFGS algorithm were compared by calculating single point DFT potential energies and by comparing structures with root-mean squared deviation (RMSD). We use term RMSE to refer prediction error in the case of testing force norm prediction with EMLM and with term RMSD we refer to the structural difference of atomic configurations. With terminology we want to distinguish that they are measuring two different kinds of differences. Here the RMSD is calculated between the final structure from the DFT level optimization and configurations of interest from ML optimization. Only moving atoms, except hydrogen atoms, are included into the RMSD calculation. 

As the ML method has always some level of uncertainty in both force norms and directions, the maximum step size might affect  the convergence. If for one element the force is overestimated, the optimization would scale all requested steps collectively letting the atom affected by the largest force be moved the most and the rest are moved just slightly. Hence, too large step size might lead to back and forth movement, when the atom with overestimated force overshoots and passes a minimum. A small step size reduces the possibility of overshooting and the BFGS approximation of Hessian matrix is updated with more modest rate than with a large maximum step size. 

\begin{figure*}
    \centering
    \includegraphics[width=\columnwidth]{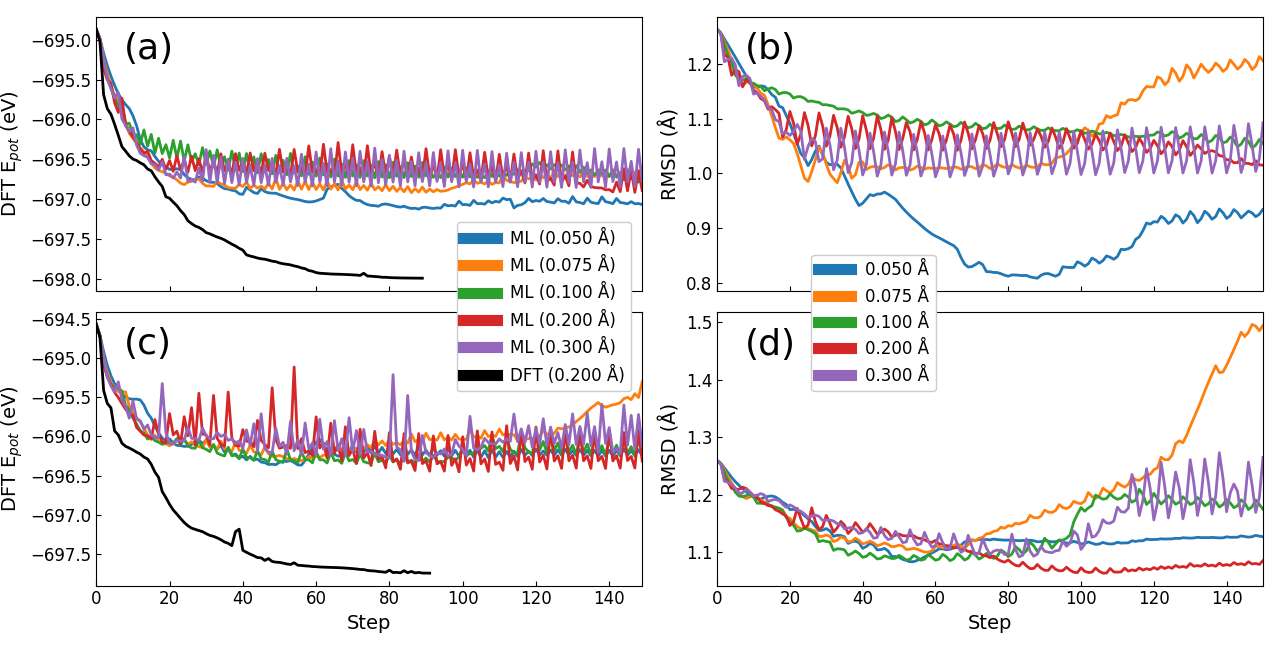}
    \caption{The evolution of the potential energy and RMSD during the optimization of \ch{Au38(SCH3)24} structures with stretched protecting units using different BFGS maximum step sizes. (a) shows the potential energy evolution for Q isomer and (b) the RMSD compared to the DFT optimized structure. (c) and (d) are corresponding plots for T isomer.}
    \label{fig:unit-epot-rmsd}
\end{figure*}

The potential energy comparison is shown in FIG. \ref{fig:unit-epot-rmsd} (a) for Q isomer and (c) for T isomer. Some differences in the convergence and the fluctuation of the potential energy are observed between different step sizes. However, all curves have converged in the similar energy level and potential energy is decreasing with a good rate. In the FIG. \ref{fig:unit-epot-rmsd} (a) maximum step size $0.05\ \text{\r{A}}$ is giving the most stable performance and it reaches the lowest energy value, even tough it is higher than what DFT optimization yields. The optimization for T isomer shows more fluctuation in FIG. \ref{fig:unit-epot-rmsd} (c) and the energy differences between DFT and ML optimizations are larger than in the case of Q isomer.

By looking at the structures and comparing them with RMSD, we can get some insight about the behavior of the ML optimization, which are not visible in potential energy. For Q isomer, the RMSD evolution in FIG. \ref{fig:unit-epot-rmsd} (b) indicates that ML optimizations with different maximum step sizes converge to somewhat different configurations. Maximum step size $0.05\  \text{\r{A}}$ manages to get closest to the DFT optimization results. This can also be seen in FIG. \ref{fig:au38struc} (c) and (e), where the structures are visualized. They have very close resemblance. 

The optimizations of the T isomers are seen to converge into very similar RMSD values in FIG. \ref{fig:unit-epot-rmsd} (d). After about 80 optimization steps the differences start to emerge. This is caused mostly by the two core gold atoms. As seen in the FIG. \ref{fig:au38struc} (d) and (f), during the ML optimization two core gold atoms are not placed as deep into the core as with DFT, which leaves protecting unit protruding from the cluster. As the convergence criterion is not reached and optimization continues, BFGS forces this unit to bend while trying to minimize the potential energy. However, even if DFT and ML optimizations lead to somewhat different structures, the potential energy is shown to be surprisingly stable.

\subsubsection{Optimization of the MD snapshots of the \ch{Au25(SCH3)18} and \ch{Au38(SCH3)24} nanoclusters}

\begin{figure}
    \centering
    \includegraphics[width=\columnwidth]{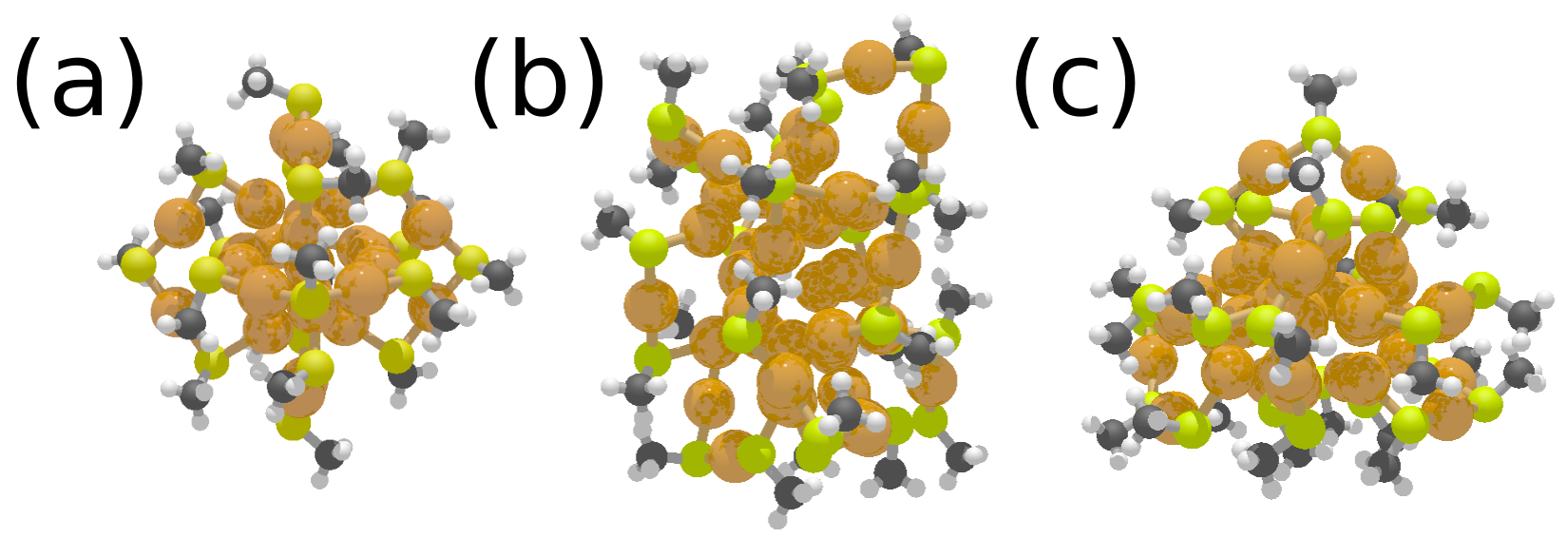}
    \caption{(a) Extracted 1500th configuration of the [\ch{Au25(SCH3)18}]$^-$ from $500\ \text{K}$ DFT MD. (b) 1000th configuration of the \ch{Au38(SCH3)24} Q isomer from the MD simulations from the reference \cite{rosalba19}. (c) 600th configuration of the \ch{Au38(SCH3)24} T isomer from the same source. Colors: orange, gold; yellow, sulfur; gray, carbon; white, hydrogen.}
    \label{fig:snapshots}
\end{figure}

The most challenging task is to optimize arbitrary configurations of the MPCs from MD simulations. The first structure to optimize is \ch{Au25(SCH3)18}. The configuration is taken from the 1500th step of the $500\ \text{K}$ MD simulation of the $[$\ch{Au25(SCH3)18}$]^-$ nanocluster with timestep of $2.0\ \text{fs}$, which is the same as used originally for MD simulations of the \ch{Au38(SCH3)24} isomers \cite{rosalba19}. The initial structure for MD was based on the experimentally found crystal structures of $[$\ch{Au25(PET)18}$]^-$ (PET= phenyl ethyl thiolate) \cite{murray08,jin08}. However, our force estimation framework does not consider the charge, therefore the optimization was done with the neutral structure. Other two test system were extracted from the MD simulations of the \ch{Au38(SCH3)24} isomers, from which the training data was collected. For Q isomer we used 1000th step and for T isomer 600th step from the original trajectories. The structures are visualized in FIG. \ref{fig:snapshots}.

First \ch{Au25(SCH3)18} was optimized with four different approaches. The first one was to simply run BFGS and optimize the whole structure. The second approach was to fix core atoms and run optimization on outer layer only. This means that unit gold atoms, sulfur, carbon and hydrogen atoms are free to move but core gold atoms are not. The last two schemes were optimizing the structure in turns, first optimizing outer layer and then the gold core. 

The most uncertain part of our ML framework is the gold core, therefore the running optimization in turns is justified. During the partwise optimization, the outer layer containing unit gold, sulfur, carbon and hydrogen atoms was optimized first 24 steps and core gold atoms were fixed. Next outer layer was fixed and core gold atoms were optimized 12 steps. This way the uncertainty inside the core does not affect directly the steps on outer layer and vice versa. The maximum step size was $0.05\ \text{\r{A}}$ as it was shown to result into stable optimizations in the previous section. 

On the other partwise optimization scheme, the uncertainty effects to the BFGS optimization were addressed. This was done by resetting the Hessian matrix approximation. Here resetting means that the Hessian matrix approximation is returned to the initial value. Optimization of MD configurations drives the ML method to its limits, therefore there is a risk that the simulations reach regions where the reliability of the method is compromised. 
This can affect the performance of the BFGS algorithm, because the usage of second order information via Hessian matrix approximation, makes it maximally affected by the noise and inaccuracies of the gradient. This is due to the ill-posedness of the noisy derivatives \cite{wang15}. Hence, readjusting the optimization might help to cope with uncertainty. The Hessian matrix approximation was reset after every 36 optimization steps (one round for both outer layer and core). 

\begin{figure}
    \centering
    \includegraphics[width=0.7\columnwidth]{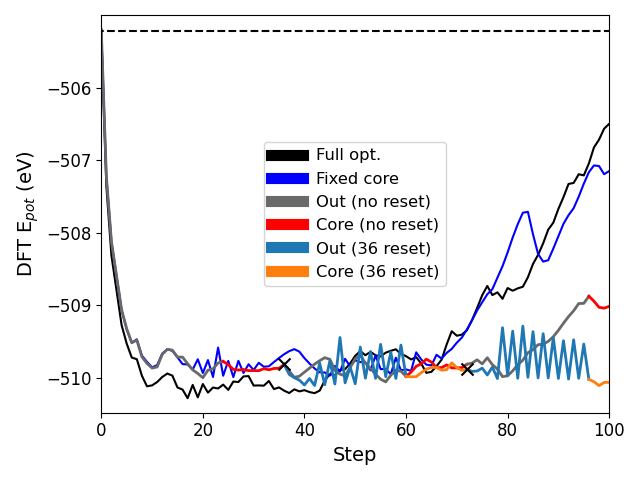}
    \caption{The evolution of single point DFT potential energies of the \ch{Au25(SCH3)18} during four different ML BFGS schemes. The potential energy of the initial configuration is highlighted with dashed line. The steps, where the Hessian matrix approximation was reset, are pointed out with black crosses on the corresponding curve.}
    \label{fig:25snap-epot}
\end{figure}

After BFGS optimization, single point DFT potential energies were calculated as before. The potential energy evolution for different optimization schemes of the \ch{Au25(SCH3)18} are shown in FIG. \ref{fig:25snap-epot}. Initially all optimization schemes perform extremely well. After about 70 steps full optimization and optimization with fixed core start to deteriorate. After 90 steps the partwise optimization without resetting of Hessian matrix approximation also shows increase in potential energy. The main cause of this phenomenon is the uncertainty building up to the Hessian matrix approximation. This conclusion is supported by the behavior of the partwise optimization, where Hessian matrix was reset. The potential energy of that optimization run is not increasing but, in contrary, is almost saturated. However, resetting introduces fluctuation to the potential energy, because the initial step of the algorithm is not as optimal as the latter ones with improved step estimation.

\begin{figure}
    \centering
    \includegraphics[width=0.5\columnwidth]{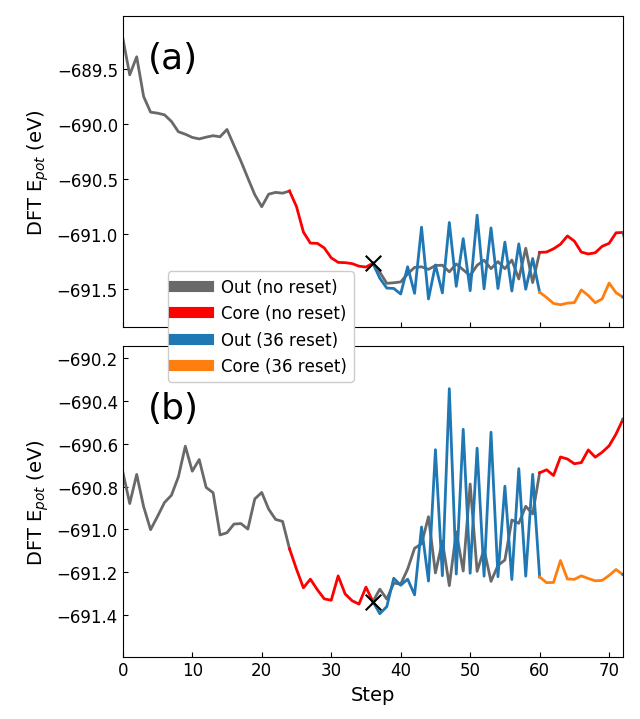}
    \caption{Evolution of DFT potential energy during the ML optimization of \ch{Au38(SCH3)24} MD snapshots for (a) Q isomer and (b) T isomer. Optimization is done in turns first optimizing 24 steps of protecting outer layer and the 12 steps of gold core. There are two different optimization approaches: normal BFGS and BFGS where Hessian matrix approximation is reset every 36 optimization step. Crosses on the curves show when the approximation of the Hessian matrix is reset.}
    \label{fig:snap-epot}
\end{figure}

The optimization of the \ch{Au25(SCH3)18} showed that partwise optimization schemes are the most promising approaches to complicated configurations. Hence, we used them to optimize previously mentioned \ch{Au38(SCH3)24} configurations. The results for the Q isomer are shown in FIG. \ref{fig:snap-epot} (a) and for the T isomer in FIG. \ref{fig:snap-epot} (b). The optimization of the Q isomer shows almost monotonous decreasing of the potential energy. However, without resetting the Hessian matrix approximation the potential energy starts to increase on the second round of the core optimization. As before, resetting improves the optimization but it introduces fluctuation to the outer layer optimization.

The optimization of the T isomer configuration is again more unstable than Q isomer as expected. The first optimization round decreased the potential energy by about $0.5\ \text{eV}$ but then the effects from the uncertainty accumulated into the Hessian matrix approximation start to emerge. This is seen as an increasing potential energy. Resetting the Hessian matrix minimizes the increase, but it fluctuations of the outer layer optimization are significant.

The results in \ref{fig:snap-epot} demonstrate the complexity of the optimization of the arbitrary \ch{Au38(SCH3)24} configurations. However, our method combining EMLMs and OAMLMs manages to decrease the potential energy by about $1.0\ \text{eV}$ for Q isomer and $0.5\ \text{eV}$ for T isomer. The \ch{Au25(SCH3)18} case is notably easier for the ML method than either of the \ch{Au38(SCH3)24} isomers. The potential energy of the system was decreased approximately by $5.0\ \text{eV}$. This is peculiar, because the model was trained with \ch{Au38(SCH3)24} and not with \ch{Au25(SCH3)18}. The most probable reason is that \ch{Au25(SCH3)18} is more well-defined than \ch{Au38(SCH3)24} isomers, therefore the simulation space is more restricted and the ML method could work on environments, which are close to its training/reference data. The optimization tests also show that the effect of uncertainty accumulated into Hessian matrix approximation could be reduced by resetting. This is valuable practical information, if one desires to use the method for real applications. Straightforward way to improve the optimization would be to add DFT-level optimization steps between the ML optimization rounds. If ML method is steered towards non-physical configurations because of the accumulated uncertainty or inputs outside the training region, the DFT optimization steps could help the overall process to converge towards a better configuration.

\section{Conclusions}

In this study we applied a novel concept of ML forces to optimize chemically complex protected \ch{Au38(SCH3)24}, \ch{Au25(SCH3)18} nanoclusters and gold-thiolate rings. The methodology was based on distance-based ML methods. The prediction of the atomic forces was divided into two parts. The prediction of the norms was done with conventional EMLM method, and the estimation of the force directions used a newly developed OAMLM method. Different parameters were tested extensively utilizing the two structural isomers of the \ch{Au38(SCH3)24} nanocluster. First we tested the performance of the model by training it with the data from one isomer and then tested it with the other. After this, another training dataset was collected using both isomers and both norm and direction prediction methods were tested.

As an application of the ML method, we used a BFGS structure optimization algorithm to utilize atomic forces estimated with EMLM and OAMLM. The optimization was first tested with gold-thiolate rings, which showed surprisingly good performance as these structures were not explicitly included in the training data. Here the method shows a great promise of generalizability. The second testing case was to optimize stretched protecting units on both isomers of the \ch{Au38(SCH3)24}. Especially the results of the isomer Q were in good agreement with the DFT. The greatest challenge was to optimize MD snapshots of \ch{Au25(SCH3)18} and \ch{Au38(SCH3)24} isomers with ML forces using different approaches to BFGS. The method performed especially well in the case of \ch{Au25(SCH3)18}, which further supports the idea of transferability, and it managed to reasonably reduce the potential energies of the \ch{Au38(SCH3)24} isomers. The tests also demonstrated that resetting of the Hessian matrix approximation is an effective approach to minimize the uncertainty effects. 

Overall, the results are promising and suggest that the method could be useful for hybrid optimization method, where coarse optimization is done with ML and fine tuning with DFT. This would also help to drive ML-based optimization away from the regions were it is uncertain. The co-operation of ML and DFT structure optimization was already briefly shown to work for gold-thiolate rings. Furthermore, the method managed to handle deformed \ch{Au25(SCH3)18} and \ch{Au38(SCH3)24} nanoclusters  with reasonable accuracy, which is an encouraging result suggesting that our methodology could be utilized on optimization of complex nanostructures.

\begin{acknowledgments}
    This work was supported by Academy of Finland through the AIPSE research program with grant 315549 to H.H. and 315550 to T.K., through the Universities Profiling Actions with grant 311877 to T.K. This work was also supported by "Antti ja Jenny Wihurin rahasto" via personal funding to A.P. ML computations were done at the FCCI node in the University of Jyv\"{a}skyl\"{a} (persistent identifier: urn:nbn:fi:research-infras-2016072533) and DFT computations at the CSC supercomputing center in Finland. We acknowledge J. Linja, J. H\"{a}m\"{a}l\"{a}inen and P. Nieminen for numerous discussions on ML methods. J. H\"{a}m\"{a}l\"{a}inen provided the basis for EMLM and RS-maximin codes.
\end{acknowledgments}

\section*{Data Availability Statement}

The Supplementary Material is available free of charge at [URL will be inserted by publisher]. It contains detailed results for the SOAP parameter testing with EMLM (figures $\text{S}1-\text{S}20$) and OAMLM (figures $\text{S}21-\text{S}24$). The method is written in Python 3.6 and it relies on Numpy \cite{numpy20}, Scikit-learn \cite{scikit-learn}, Atomic Simulation Environment \cite{ase17}, DScribe \cite{dscribe} and Scipy \cite{scipy20} packages. The parallelization of the testing and training of the methods and the BFGS optimization are done via mpi4py package \cite{mpi4py-2005,mpi4py-2008,mpi4py-2011,mpi4py-2021}. The code, optimization data and complete parameter test visualizations are available at Gitlab \url{https://gitlab.jyu.fi/aneepihl/oamlm_forces.git}.

\section*{Author declarations}
\subsection*{Conflict of Interest}

The authors declare no competing financial interest.


%

\end{document}